# Nitrogen and argon doping of niobium for superconducting radio frequency cavities: a pathway to highly efficient accelerating structures

A. Grassellino[1*], A. Romanenko[1], D. Sergatskov[1], O. Melnychuk[1], Y. Trenikhina[2], A. Crawford[1], A. Rowe[1], M. Wong[1], T. Khabiboulline[1], F. Barkov[1]
[1]Fermi National Accelerator Laboratory, Batavia, Illinois, 60510, USA
[2]Illinois Institute of Technology, 3300 S Federal St, Chicago, IL 60616

We report a surface treatment that systematically improves the quality factor of niobium radio frequency cavities beyond the expected limit for niobium. A combination of annealing in a partial pressure of nitrogen or argon gas and subsequent electropolishing of the niobium cavity surface leads to unprecedented low values of the microwave surface resistance, and an improvement in the efficiency of the accelerating structures up to a factor of 3, reducing the cryogenic load of superconducting cavities for both pulsed and continuous duty cycles. The field dependence of the surface resistance is reversed compared to standardly treated niobium.

Superconducting Radio Frequency (SRF) is a key enabling technology for essentially all new high energy and high beam power accelerators envisioned worldwide. Applications of SRF particle accelerators range from particle and nuclear physics to medicine, defense, homeland security, and industry. The typically employed cavity surface processing for achieving the state-of-the-art RF performance includes a combination of chemical treatments like electropolishing (EP), buffered chemical polishing (BCP) and heat treatments [1]. Decades of SRF R&D in laboratories and universities have enabled SRF niobium cavities to perform reliably and systematically with large quality factors Q (ratio of cavity stored energy and dissipated power per RF cycle) typically determined by the adopted surface processing, up to accelerating gradients $E_{acc} \sim 40$ MV/m corresponding to peak magnetic fields on the cavity surface of $B_{pk} \sim 160\text{-}180$ mT. However, SRF niobium cavities still suffer of a decreased efficiency with the increasing RF accelerating voltage $E_{acc}$. The phenomenon, known as medium field Q-slope [1-5] (MFQS) consists of a degradation of the cavities quality factor up to 50% at peak magnetic fields ~ 60-80 mT, which correspond to the operating field range of all the continuous wave present and future accelerators, like for example Project X [6], NGLS [7] and ERLs [8]. Eliminating this Q degradation is a crucial issue for continuous wave (CW) accelerators, since it would contribute to cut significantly capital and operating costs. The quality factor of a SRF cavity is inversely proportional to the niobium RF surface resistance, consisting of two components: the temperature dependent Bardeen Cooper Schrieffer (BCS) part, due to thermal excitation of quasi-particles at a finite temperature, and a temperature independent component defined as the residual resistance due to several factors like for example trapped magnetic flux, normal conducting precipitates etc. The key to reduce the large costs associated with cryogenic losses in SRF accelerators is to find a surface treatment that can minimize the two surface resistance components and their field dependence. The origin and the nature of the field dependence of surface resistance in

---

niobium cavities remain poorly understood, despite having been the subject of advanced investigations for many years [1, 9, 10, 11, 12]. Also, while an empirical cure for the high field losses – above 80 mT peak magnetic fields- has been identified in the '120°C bake' [13, 14, 15], up to today there is no known surface processing, which reproducibly eliminates the medium field RF losses. Only two separate experimental results have been reported where the field dependence of the surface resistance is actually reversed [16, 17], but neither the origin of, nor the empirical process leading to the Q field dependence reversal for those two cases have been understood, and results have not been reproduced. In this letter we report the finding of a surface treatment that reproducibly reverses the field dependence of the surface resistance of niobium cavities, in particular of its temperature dependent component. The treatment produces also lower than typical residual resistances, leading overall to quality factors two to three times higher than those obtained with standard treatments.

The studies presented began with the idea to lower the SRF niobium cavities surface resistance by realizing a RF layer of a superconducting compound with a higher critical temperature than niobium. For years a large R&D effort at several institutions and universities has been ongoing towards the realization of materials alternative to niobium, like $Nb_3Sn$, $NbN$, $NbTiN$ for SRF cavities via deposition of thin films or via bulk diffusion [17, 18, 19, 20, 21, 22]. Several results have been reported where a cavity surface layer possessing higher critical temperature has been obtained [23, 24, 25, 26] leading to extremely low BCS surface resistance values. However, a typical issue limiting the performance of such coated cavities is the presence of large residual resistances of ~ hundreds of nΩ, possibly due to the co-formation of unwanted phases with very low critical temperatures. To date, only one result has been obtained on a $Nb_3Sn$ cavity made at University of Wuppertal with both extremely low BCS and residual resistances, but limited however to low accelerating gradients due to a strong Q degradation with field [27].

Table 1. List of SRF niobium cavities used in the study and respective parameters and performance post nitrogen heat treatment, for different amount of material removal via EP.

| CAVITY ID | Type | Treatment | Subsequent cumulative material removal via EP for each RF test [μm] | Highest Q measured at T=2K (correspondent to material removal in bold); max Q value located at ~ $B_{pk}$ [mT] |
|---|---|---|---|---|
| TE1AES016 | Large grain EP | 1000°C 1 hour with ~ $2\times10^{-2}$ Torr p.p. nitrogen | **80** | $(7.4\pm1.4)\times10^{10}$, 40 mT |
| TE1AES003 | Fine grain BCP | 1000°C 10 min with ~ $2\times10^{-2}$ Torr p.p. nitrogen | **10**, 60 | $(4.1\pm0.6)\times10^{10}$, 50 mT |
| TE1AES005 | Fine grain EP | 1000°C 1 hour with ~ $2\times10^{-2}$ Torr p.p. nitrogen | 20, 40, **80** | $(4.2\pm0.13)\times10^{10}$, 70 mT |
| TE1NR005 | Fine grain EP | 800°C 3 hours in UHV, followed by 800°C 10 min with ~ $2\times10^{-2}$ Torr p.p. nitrogen | **5**, 15 | $(5.3\pm0.85)\times10^{10}$, 70 mT |

We decided to investigate the possibility of forming a niobium nitride layer by reacting bulk niobium cavities with nitrogen in a high temperature UHV furnace. Three single cell 1.3 GHz TESLA shape fine grain cavities and one large grain cavity were baked in the high temperature furnace with nitrogen gas injection for a certain period of time, as summarized in Table 1. A description of the furnace used in these studies can be found in [29]. The cavities then received the typical preparation for RF testing, consisting of a high-pressure water rinse (HPR) [1]. The RF characterization of the cavities consists of measurements of the quality factor versus the accelerating gradient via standard phase lock techniques, as described in [30]. We used fixed input antennas which were trimmed to a length matching the cavity Q at T=2 K with the external quality factors $Q_{ext}$ in the range $5\times10^9 < Q_{ext} < 8\times10^{10}$. Corresponding case-by-case calculated errors on the Q values are reported in Table 1. Directly after nitrogen treatment, all cavities at T = 2 K showed quality factors in the range ~ $10^7$, very poor compared to the routinely achieved Q values for standardly treated niobium cavities at this frequency and temperature of ~ $2\times10^{10}$. These results seemed to indicate formation of unwanted poorly superconducting NbN phases. The cavities were then subject to different amounts of material removal of the niobium RF surface via EP, as shown in Table 1 respectively, after which they received again HPR and full RF measurements were repeated. Surprisingly, after a certain amount of material removal via EP post nitrogen treatment, the performance of all the nitrogen treated cavities were drastically improved rather than being back to the standard performance which are routinely obtained with an EP surface. Fig. 1 shows the results of the RF measurements, shown as the cavities Q as a function of the accelerating field in the cavity. The $Q(E_{acc})$ curves at T = 2 K for the four cavities treated with nitrogen plus some amount of EP are compared to a typical $Q(E_{acc})$ curve for an electropolished cavity.

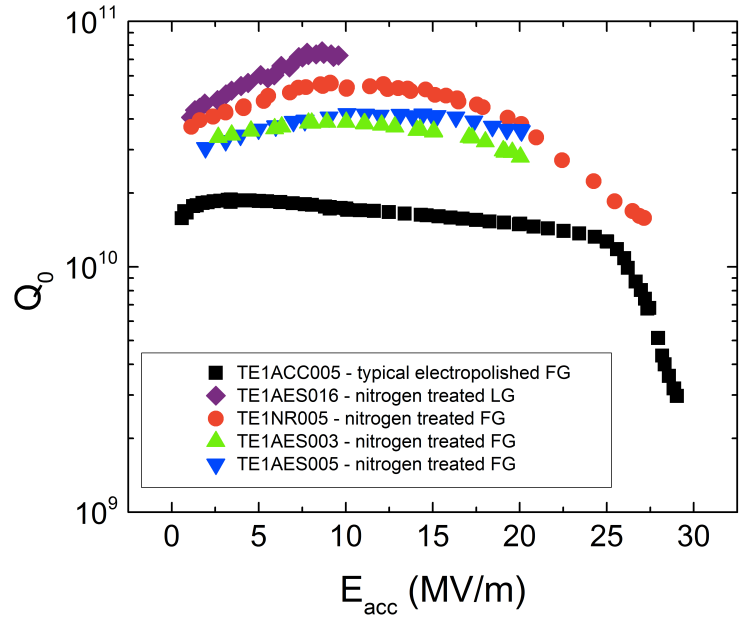

Figure 1. Comparison of quality factors versus accelerating field for nitrogen treated cavities and standardly adopted treatment (electropolishing). Improvements in quality factors up to a factor of 3.5 are found, in the region of accelerating gradients of interest for CW SCRF accelerators.

Two main things in these curves are striking: 1) the measured Q values at T = 2 K are significantly higher than typical values for standardly treated surfaces; 2) the field dependence of the quality factor is reversed, showing an extended anti-Q-slope, contrary to the standard medium Q-slope behavior of niobium cavities. The medium field Q-slope in Nb cavities after standard surface treatments is due to the increase in both residual resistance and BCS resistance with the amplitude of the RF field, as it has been recently shown in [31]. To gain a better understanding of the origin of the Q improvement in the nitrogen treated cavities, we adopted the same technique for decoupling residual resistance and BCS resistance as a function of field as described in [31]. The results are shown in Fig. 2 in comparison to the standard cavity surface treatments: EP, BCP, before and after the 120°C bake. Such an analysis reveals that the nitrogen treatment decreases the BCS resistance at low RF fields to values comparable to those obtained with the 120°C bake. However, as the field increases, the BCS resistance increases dramatically for the 120°C case, whereas it actually decreases even further for the nitrogen treatment case. This strong difference leads to a BCS resistance ~2.5 times lower compared to the 120°C treatment and ~3.5 times lower compared to standard EP or BCP. An unusual and extended anti Q-slope therefore stems from this inverted field dependence of the BCS resistance. All four cavities were also associated with lower than typical residual resistance values, as shown for example in Fig. 2.

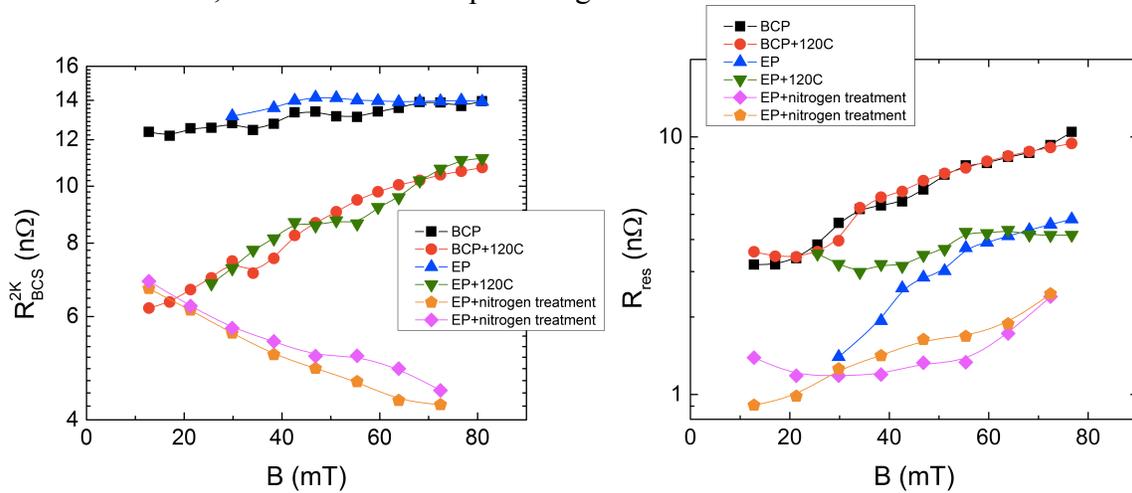

Figure 2. Temperature-dependent (BCS) and temperature independent (residual) components of the microwave surface resistance. A comparison is shown for the nitrogen treated cavities and those after standard treatments including EP, BCP and 120°C bake.

Additional studies were performed on the microwave surface resistance evolution as a function of material removal post nitrogen treatment. In Fig. 3 we show the evolution of the $Q(E_{acc})$ curves of the three different fine grain cavities for different amount of material removal via EP. For TE1AES005 it is revealing to notice how the Q-curve changes: for 20 and 40 μm removal the low field Q is high compared to that of standard EP surfaces, but the medium field Q-slope still dominates the curve. However, with further 40 μm removal (80 μm total removal post nitrogen), an extended anti Q-slope appears, indicating that there is an optimal amount of material removal post nitrogen treatment, which gives the high Q performance up to the higher fields. For the cavity TE1NR005 the anti- Q-slope is already found with just 5 μm removal, and interestingly further 10 μm EP 'straightens' the Q curve: the curve is now 'in between' the anti-Q-slope and the standard medium field Q-slope behavior. Q(T=2 K) is still atypically high at ~ $3\times10^{10}$, 17 MV/m, plus no medium field Q-slope and no high field Q-slope are observed up to the quench field of ~ 27 MV/m. Similar behavior is observed for TE1AES003, for which 10 μm removal gives the extended anti-slope, and 60 μm EP starts to show some mild MFQS, although performance is still somewhat in between those of regular EP surfaces and the anti-slope. The observations on the four different cavities are consistent: in all cases, the highest Q and anti Q-slope are found at about a quarter of the nitrogen diffusion length in niobium, calculated for each cavity bake cycle parameters (temperature and length of nitrogen exposure). The performances revert back to those of standard cavities towards the end of the nitrogen diffusion tail. Further studies of the performance post nitrogen treatment at finer EP steps are ongoing, along with investigations for cavities baked at different temperature, for different time length and for different partial pressures of nitrogen.

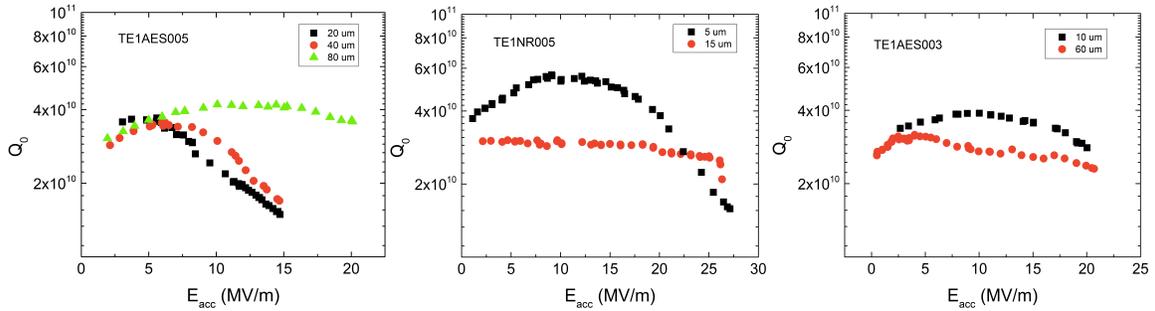

Figure 3. Performance of the three fine grain nitrogen treated cavities for different amount of material removal post nitrogen treatment.

The results presented seem to indicate the existence of an optimal nitrogen to niobium concentration ratio, which alters beneficially the cavity microwave surface resistance. There are two possible scenarios: 1) formation of niobium nitride phases with higher critical temperature than niobium; 2) nitrogen as an interstitial in the niobium lattice. A $T_c$ measurement was performed for cavity TE1AES016 resulting in the standard niobium critical temperature of ~ 9.2 K. This rules out that a continuous layer of niobium nitride covers the cavity surface; however it does not exclude the possibility that islands of NbN could form only at some nucleation centers and coexist with the standard niobium surface. SEM images of NbN formed via bulk diffusion in [24] show how the high $T_c$ phases (δ, γ) can form as the most internal layers, while the most external layers can be the poorly superconducting hexagonal phases. This could correlate with our findings where the cavity Q is poor right after the nitrogen treatment, and it becomes exceptionally high after a certain amount of material removal via EP. Samples studies are ongoing utilizing XRD, XPS and Auger spectroscopy to investigate the underlying processes. Preliminary XPS results indicate average surface concentrations of nitrogen in the range of 15-25 atomic %, decaying to ~ 6-10% at depths comparable to the amount of material removal via EP which leads to the high Q performance. Images collected via laser confocal microscopy confirm the formation of precipitates at the surface- forming differently for different grains- and precipitates are still found after several microns of ion sputtering. However, while some NbN precipitates certainly form at the surface, we have found no evidence that they remain after the amount of material removal comparable to the EP done on the cavities. XPS and XRD measurements performed on a sample treated for 10 minutes at 800°C in the partial pressure of nitrogen ~ $10^{-2}$ Torr followed by 5 micron EP, have not identified presence of nitride phases. Even though further studies are needed, this favors the hypothesis of performance improvement due to the presence of a certain concentration of nitrogen as an interstitial in the Nb lattice. In this case, a possible physical mechanism could be that nitrogen in niobium as interstitial can act as a trap for hydrogen. It has been shown that due to the trapping process the hydrogen-induced resistivity increase is reduced and the occurrence of precipitation is shifted to higher hydrogen concentrations [32]. This mechanism is consistent with findings in [33] and a recent RF losses model based on nanohydride precipitates [34]. Both possible scenarios have important consequences: if the performance improvement comes from the presence of NbN islands, then these results would demonstrate that a material with higher critical temperature than niobium can lead to higher quality factors at higher gradients than measured before for bulk Nb cavities, and that the lower critical field of the material does not represent a limit for the RF performance. If instead the improved performances stem from nitrogen as interstitial, which for example might neutralize hydrogen and prevent formation of lossy nano-hydrides, then the implications might be even more profound: it might mean that what for decades was believed to be the theoretical limit for niobium microwave surface resistance, it was not. Some recent theoretical calculations predict inverse RF field dependence of the Mattis-Bardeen surface resistance [35], and preliminary fits based on this model are in good agreement with the presented results.

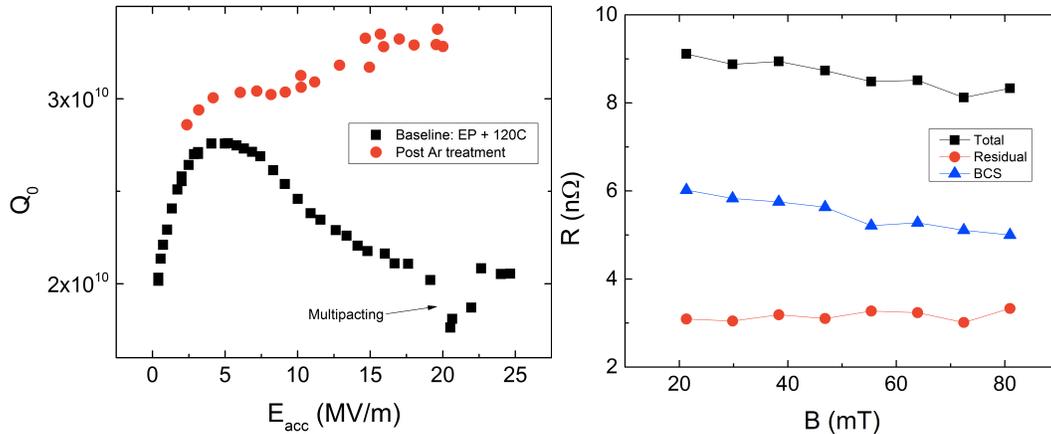

Figure 4. Left: Quality factor versus accelerating field at 2 K for TE1CAT003, comparison for baseline treatment and after the high temperature bake in argon atmosphere. Right: deconvolution of TE1CAT003 surface resistance in the BCS and residual part versus peak magnetic fields.

We performed an additional experiment to investigate if the improvement in surface resistance could originate from the presence of interstitial nitrogen and not from a compound with higher critical temperature. We baked another fine grain 1.3 GHz cavity (TE1CAT003) in the same furnace used for the nitrogen treated cavities but this time diffusing an inert gas into niobium; the cavity was baked at 800°C in UHV for an hour, followed by injection of ~$2\times10^{-2}$ Torr partial pressure of argon for one hour at 800°C, then received high-pressure water rinse and RF testing. The resulting Q measured at 2 K was extremely poor, of the order of $10^8$. The cavity then received approximately 5 μm removal via electropolishing, HPR and again RF testing. Interestingly, results showed again an extended anti-Q-slope up to $B_{peak}$ ~ 90 mT, leading to an atypically high Q ~ $3.5 \times 10^{10}$ at 90 mT, T=2K. The comparison of the $Q(E_{acc})$ curves before and after the argon treatment in Fig. 4 clearly shows the effect of reversal of the medium field Q-slope. Deconvolution of the field dependence of residual and BCS resistance reveals again the atypical BCS component decreasing with field while the residual resistance remains constant up to the highest reached fields. Since argon is an inert gas, it can only be in the lattice as an interstitial without forming any compound. Therefore, these results seem to suggest that there is a crucial role of impurity doping in the surface resistance improvement of SRF niobium cavities that we have found. It is interesting to notice that in the single anti-Q-slope result reported in [17] argon was also injected in the furnace during the cooldown from 1400°C to room temperature, and that witness samples analysis revealed significant concentration of other surface impurities like titanium and nitrogen. Therefore impurity doping (nitrogen, argon, titanium) could be the common reason of the anti-Q-slope cavity results. Our findings suggest that introducing a small amount of impurities at the surface might need to become a standard practice for cavity surface preparation to realize the niobium technology at the full potential. Ideally, impurity doping should be limited only to the RF layer to minimize its effect on thermal conductivity so that even higher gradients could be systematically achieved.

In summary, we have reported the experimental findings of a surface treatment for SRF niobium cavities: a combination of vacuum heat treatment in nitrogen or argon atmosphere and a subsequent material removal by electropolishing produces unprecedented low values of the microwave surface resistance. The described treatments reverse the medium field Q-slope due to the decreasing with field Mattis-Bardeen surface resistance and, unlike after all standard processing techniques, the cavity quality factor grows with the increasing amplitude of the RF field level in the cavity. These findings open the way for up to a factor of 3 cut in cryogenic losses in CW SRF accelerators, and offer a long sought solution to the medium field Q-slope problem in SRF cavities.

*Acknowledgements*
The authors would like to thank Dr C. Ginsburg, Dr L.D. Cooley, Dr V. Yakovlev and Dr R. D. Kephart for supporting this R&D work, for several insightful discussions and for providing feedback on the manuscript. We would like to thank also the technical support of the Fermilab cavity processing team, and Fermilab T&I department for the support with cavity testing. An acknowledgment for several insightful discussions on the results goes to Prof. Enzo Palmieri, and for his extensive work on NbN that originally inspired these studies. Fermilab is operated by Fermi Research Alliance, LLC under Contract No. DE-AC02-07CH11359 with the United States Department of Energy. A. R. and Y. T. are partially supported by the DOE Office of Nuclear Physics.